
\magnification=\magstep1
\overfullrule=0pt\nopagenumbers
\hsize=15.4truecm
\line{\hfil UQAM-PHE-94/01}
\line{\hfil hep-ph/9403297}
\centerline{\bf QCD CORRECTIONS TO THE $t\rightarrow H^+b$\ DECAY WITHIN}
\centerline{\bf THE MINIMAL SUPERSYMMETRIC STANDARD MODEL}
\vskip1cm
\centerline{HEINZ K\"ONIG\footnote*{email:konig@osiris.phy.uqam.ca}}
\centerline{D\'epartement de Physique}
\centerline{Universit\'e du Qu\'ebec \`a Montr\'eal}
\centerline{C.P. 8888, Succ. Centre Ville, Montr\'eal}
\centerline{Qu\'ebec, Canada H3C 3P8}
\vskip2cm
\centerline{\bf ABSTRACT}\vskip.2cm\indent
I present the contribution of gluinos and scalar quarks to
the decay rate of the top quark into a charged Higgs boson and
a bottom quark within the minimal supersymmetric standard
model, including the mixing of the scalar partners of
the left- and right-handed
 top quark. I show that for certain values of the
supersymmetric parameters the standard QCD loop corrections
to this decay mode are diminished or enhanced by several
10 per cent. I show that not only a small value of 3 GeV for the
gluino mass (small mass window) but also much larger
values of several hundreds of GeV's
have a non-neglible effect
on this decay rate, against general belief.
 Last but not least, if the ratio of the
vacuum expectation values of the Higgs bosons are taken
in the limit of $v_1\ll v_2$ I obtain a drastic
enhancement due to a $\tan\beta$\ dependence in the couplings.
\vskip2cm
\centerline{ March 1994}
\vfill\break
{\bf I. INTRODUCTION}\hfill\break\vskip.12cm\noindent
Recently there have been a lot of interest in the
electroweak loop corrections [1,2]
as well as to the QCD loop corrections [3,4,5,6]
to the top quark decay into a charged Higgs boson
and a bottom quark.\hfill\break\indent
In the standard model we have no charged Higgs particle
and therefore this decay can be used as a test for models
beyond the standard model; such as a two Higgs
doublet model [see eg.
7 and references therein] and the minimal supersymmetric
extensions of the standard model (MSSM) [8,9], which is the
 favorite model beyond the standard model.
\hfill\break\indent
In this paper I take the last one as the underlying
model to consider the QCD corrections to the $t\rightarrow H^+b$\
decay mode. In [6] it was shown that the effect
of the mass of the bottom quark to this decay rate is negligible.
It was also shown that the ratio of the first
order to the zeroth order is constant at about $-9\%$ for a
wide range of the Higgs mass ($0\le m_{H^+}\le 90$\ GeV); the
top quark mass was taken to be $150$\ GeV.\hfill\break\indent
In this paper I present the QCD loop corrections to the
 $t\rightarrow H^+b$\ decay if gluinos and scalar quarks
are taken within the loop as shown in Fig.1. Throughout
the calculations I neglect the mass of the bottom quark,
but I do not neglect the mixing of the scalar partners of
the left- and right-handed
top quark, which is proportional to the top quark
mass.\hfill\break\indent
In the next section I present the calculation of the Feynman
diagram given in Fig.1 and give the results. In the last
section I discuss the results and end with the conclusions.
\hfill\break\vskip.12cm\noindent
 {\bf II. SUSY QCD CORRECTIONS
}\vskip.12cm
In the MSSM the interaction Lagrangian relevant to the
decay $t\rightarrow H^+b$\ is given by:
$${\cal L}={{g_2}\over{\sqrt{2}}}{{m_{\rm top}}\over{m_W}}V_{tb}
\cot\beta
\bigl (H^+\overline t P_Lb+H^{+*}\overline bP_R t\bigr )\eqno(1)$$
where $H^+=\cos\beta H_2^+-\sin\beta H_1^+$,
$\cot\beta=v_1/v_2$\ is the ratio of the vacuum expectation
values (vev) of the two Higgs doublets
and $P_{L,R}$\ are the chiral projection operators.\hfill\break\indent
The Lagrangian of eq.(1) leads to the following decay rate
for $m_b\ll m_{\rm top}$:
$$\Gamma^0(t\rightarrow H^+b)={{G_F}\over{\sqrt{2}}}\vert
V_{tb}\vert^2\cot^2\beta{1\over{8\pi}}m_{\rm top}^3\bigl (1-
{{m_{H^+}^2}\over{m^2_{\rm top}}}\bigr)^2\eqno(2)$$
To calculate the 1 loop diagram given in Fig.1 we need the
couplings of the scalar quarks to the charged Higgs boson
and the scalar-quark-gluino-quark coupling. The first coupling
is given in Fig.115\footnote*{$\mu$\ has to be replaced by $-\mu$}
in [7] and the latter one in eq.(C89) in [9].\hfill\break\indent
When neglecting the bottom quark mass only the scalar partner of
the left handed bottom quark $\tilde b_L$\ occurs within the loop whereas
for the top quark we have to take both left- and right-handed
superpartner $\tilde t_L$\ and $\tilde t_R$\ into account.
Furthermore since the mixing of $\tilde t_L$\ and $\tilde t_R$\
is proportional to the top quark mass we have to include the
full scalar top quark matrix which is given by [10]:
$$M^2_{\tilde t}=\left(\matrix{m^2_{\tilde t_L}
+m_{\rm top}^2+0.35D_Z^2&
-m_{\rm top}(A_{\rm top}+\mu\cot\beta)\cr-m_{\rm top}
(A_{\rm top}+\mu\cot\beta)&
m^2_{\tilde t_R}+m^2_{\rm top}+0.16D_Z^2\cr}\right)\eqno(3)$$
where $D_Z^2=m_Z^2\cos 2\beta$.\hfill\break\indent
$m^2_{\tilde t_{L,R}}$\ are soft breaking masses.
 $A_{\rm top}$\ is a trilinear
scalar interaction parameter and $\mu$\ is the supersymmetric
mass mixing term of the Higgs bosons. The mass eigenstates
$\tilde t_1$\ and $\tilde t_2$\ are related to the current
eigenstates $\tilde t_L$\ and $\tilde t_R$ by:
$$\tilde t_1=cos\Theta\tilde t_L+\sin\Theta\tilde t_R\qquad
\tilde t_2=-\sin\Theta\tilde t_L+\cos\Theta\tilde t_R\eqno(4)$$
 In the following we take
$m_{\tilde t_L}=m_{\tilde t_R}=m_S=A_{\rm top}$\
(global SUSY), $m^2_{\tilde b_1}=m_S^2
-0.42D^2_Z$\ and $m^2_{\tilde b_2}=m_S^2
-0.08D^2_Z$.
With negelcting bottom quark mass
the scalar partners of the left and right handed bottom quarks
do not mix and therefore $m_{\tilde b_L}=
m_{\tilde b_1}$.
The gluino mass $m_{\tilde g}$\ is a free parameter,
which in general is supposed to be larger than 100 GeV, although
there is still the possibility of a small gluino mass window in
the order of 1 GeV [11,12].\hfill\break\indent
The results of the calculation of the loop diagram in Fig.1
are finite, there are no dimensional divergencies. As a result
I get for the first order in $\alpha_s$:
$$\eqalignno{\Gamma^1_{MS}(t\rightarrow H^+b)=&\Gamma^0
(t\rightarrow H^+b)\Bigl\lbrack 1-{{2\alpha_s}\over{3\pi}}
(S+A)\Bigr\rbrack&(5)\cr
S=&S_t+{{m_{\tilde g}}\over{m_{\rm top}}}S_{\tilde g}\cr
A=&A_t+{{m_{\tilde g}}\over{m_{\rm top}}}A_{\tilde g}\cr
S_t=&K_{11}\lbrack c^2_\Theta C_1^{\tilde b_1\tilde t_1}
+s^2_\Theta C_1^{\tilde b_1\tilde t_2}\rbrack
+K_{21}\lbrack s_\Theta c_\Theta(C_1^{\tilde b_1\tilde t_1}
-C_1^{\tilde b_1\tilde t_2})\rbrack\cr
A_t=&S_t\cr
S_{\tilde g}=&K_{11}\lbrack c_\Theta s_\Theta
(C_0^{\tilde b_1\tilde t_2}-
C_0^{\tilde b_1\tilde t_1})\rbrack-K_{21}\lbrack c_\Theta^2
C_0^{\tilde b_1\tilde t_2}
+s^2_\Theta C_0^{\tilde b_1\tilde t_1}\rbrack\cr
A_{\tilde g}=&S_{\tilde g}\cr
K_{11}=&1-{{m^2_W}\over{m^2_{\rm top}}}\tan\beta\sin 2 \beta \cr
K_{21}=&{1\over{ m_{\rm top}}}(A_{\rm top}+\mu\tan\beta)\cr
C_0^{\tilde b_j\tilde t_i}=&-\int\limits_0^1d\alpha_1
\int\limits_0^{1-\alpha_1}d\alpha_2
{{m^2_{\rm top}}\over {f_{\tilde g}
^{\tilde b_j\tilde t_i}}}\cr
C_1^{\tilde b_j\tilde t_i}=&-\int\limits_0^1d\alpha_1
\int\limits_0^{1-\alpha_1}d\alpha_2
{{m^2_{\rm top}\alpha_1}\over{ f_{\tilde g}
^{\tilde b_j\tilde t_i}}}\cr
f_{\tilde g}^{\tilde b_j\tilde t_i}=&m^2_{\tilde g}-(m^2_{\tilde g}-
m_{\tilde t_i}^2)\alpha_1-(m^2_{\tilde g}-m^2_{\tilde b_j})\alpha_2-
m^2_{\rm top}\alpha_1(1-\alpha_1-\alpha_2)
-m^2_{H^+}\alpha_1\alpha_2\cr}
$$
where $c_\Theta=\cos\Theta$\ and $s_\Theta=\sin\Theta$.
The S and A in eq.(5) indicate that the contribution
comes from the scalar - and axial scalar coupling
of the matrix element.
The Feynman integration can be done numerically.
\hfill\break\indent
In eq.(10) in [6] the authors present the results of the
standard QCD 1 loop corrections within the two Higgs doublet
model and the MSSM, which I will include in my calculation.
As a final result I obtain:
$$\eqalignno{\Gamma^1(t\rightarrow H^+b)=&\Gamma^0(t\rightarrow
H^+b)\Bigl\lbrack 1+{{4\alpha_s}\over{3\pi}}\tilde G'_+-
{{2\alpha_s}\over{3\pi}}(S+A)\Bigr\rbrack\cr
\tilde G'_+=&2{\rm Li}_2(1-\chi^2)-{{\chi^2}\over{1-\chi^2}}
\log{\chi^2}+\log{\chi^2}\log{(1-\chi^2)}+{1\over{\chi^2}}
(1-{5\over 2}\chi^2)\cdot\cr
&\log{(1-\chi^2)}
-{{2\pi^2}\over{3}}+{9\over 4}\cr
\chi^2=&{{m_{H^+}^2}\over{m_{\rm top}^2}}&(6)\cr}$$
Eq.(6) gives the full ${\cal O}(\alpha_s)$\ QCD correction
within the MSSM. In the next section I will discuss
the result.
\hfill\break\vskip.12cm\noindent
{\bf III. DISCUSSIONS}\vskip.12cm
To compare the standard QCD correction given in [6] with the
gluino contribution via eq.(5)
I present in Fig.2--4
the results for different masses of the gluinos
and different values of the $\mu,\ A_{\rm top}$\ parameters
and $\tan\beta$.
In the MSSM we have $m_{H^+}^2=m_W^2+m_{H_3^0}^2$\ where
$H^0_3$\ is the pseudo Higgs particle. That is the mass of the
charged Higgs particle has to be larger than the mass of the
W boson.
I have set the
top quark mass to be the recently released CDF value of 174 GeV
[13], the charged Higgs mass to be equal $m_W$\ and
the vevs to be equal $v_1=v_2$. In Fig.2 I set $\mu=0=A_{\rm top}$.
It should be kept in mind that $\mu=0$\
is unrealistic, because it leads in general to chargino masses
below the experimental limit of 45 GeV. With $\sin\Theta=0=
K_{21}$\
 we see from eq.(5) that only $S_t$\ and $A_t$\
lead to a contribution to the first order in $\alpha_s$.
The scalar top quark masses vary from 181 GeV to 482 GeV
and $m_{\tilde b_i}=m_S$.
In Fig.2 I present the results for three different values
of the gluino mass that is 3 GeV (solid line), 100 GeV (dotted
line) and 500 GeV (dash-dotted line). The standard contribution
of [6] is presented by the solid straight line and lies at
$-9.5$\%. \hfill\break\indent
As a result we have that for small values of the scalar mass
$m_S$\ and a small gluino mass the standard result is diminished
by a non-negligible amount down to $-7$\%. If the Higgs mass
is enhanced all curves are pushed up closer to 0, but the
shape of the curves remain the same. For $m_{H^+}=120$\ GeV
the standard QCD correction is about $-8.1$\%. The effect of
$v_1\ll v_2$\ is that the curves for the different gluino
masses are pushed closer to the standard QCD correction.
\hfill\break\indent
In Fig.3 we consider the case with $\mu=500$\ GeV and
$A_{\rm top}=m_S$\ again with $v_1=v_2$\ and the same three different
gluino masses. In this case the lighter scalar top quark
mass is about 250 GeV for $m_S$\ smaller than 100 GeV, decreases
constantly to about 70 GeV for $m_S=350$\ GeV and increases
again to 260 GeV in the range considered here.
The heavier one varies from
358 GeV to 631 GeV. Here
$\cos\Theta=1/\sqrt{2}$\ and $K_{21}> K_{11}$.
Therefore in this case $S_{\tilde g}$\ and $A_{\tilde g}$\
contribute more to the decay rate than $S_t$\ and $A_t$.
\hfill\break\indent
As a result we have in this case that the standard QCD corrections
are diminished for small gluino masses whereas we get an
enhancement up to $-18$\% for a gluino mass of 500 GeV.
Changing the $\mu$- parameter hardly effects the results.
Enhancing the Higgs mass leads to the same changings as
mentioned above.\hfill\break\indent
We obtain a totally different result in
this case when we consider $v_1\ll v_2$. From eq.(5) we
see that the coupling $K_{21}$\ is dominated by the $\tan\beta$\
term and therefore we have $K_{21}\gg K_{11}$. The gluino mass
becomes more important. For very large gluino masses ($m_{\tilde g}
\gg 100$\ GeV) the 1 loop contribution $\Gamma^1(t\rightarrow
H^+b)$\ is decreasing again.
In Fig.4 we have taken e.g. $v_2=10\cdot v_1$\ with $\mu=500$\ GeV
and $A_{\rm top}=m_S$\ as before. Here $\cos\Theta\approx 1/\sqrt{2}$,
the lighter scalar top quark mass is about 115 to 110 GeV for
$m_S$\ smaller than 100 GeV and increases constantly to 379 GeV
for $m_S=450$\ GeV.
The heavier one varies from 219 GeV to 564 GeV.
The heavy scalar bottom quark mass varies from 78 GeV
to 454 GeV and the lighter one from 56 GeV to 451 GeV.
As a result we see that the gluino mass contribution
enhances the standard QCD correction drastically.
This decay
mode therefore can be used to put constraints on the ratio
of the vevs $v_1$\ and $v_2$.
Smaller values for $\mu$\ diminishes the results whereas
higher values for $\mu$\ enlarges them.\hfill\break\indent
$v_1\ll v_2$\ has to be taken with care, because we neglected
the mass of the bottom quark. If I take $v_2=2\cdot v_1$\ in the
last case, that is still in the limit of $m_b\tan\beta\ll
m_{top}\cot\beta$ I do get the same shape as in Fig.4, but
with the values pushed a bit closer to the standard model.
\hfill\break\vskip.12cm\noindent
{\bf III. CONCLUSIONS}\vskip.12cm
In this paper I presented the results of the calculation of
the 1 loop correction to the decay rate $t\rightarrow H^+b$\
when gluinos, the scalar partner of the left- handed bottom
quark and the scalar partners of the left- and right- handed
top quark are taken within the relevant loop diagram.
We presented two cases with the vevs of the Higgs bosons to
be taken equal where $\mu=0=A_{\rm top}$\ and the other one with
$\mu=500$\ GeV and $A_{\rm top}=m_S$. I have shown that the standard
QCD corrections are changed from $-9.5$\% to $-7$\%
for a small gluino mass and to $-18$\% for a large gluino
mass if we set the top quark mass to be 174 GeV and the
charged Higgs boson mass to be equal to $m_W$. Enhancing the Higgs
mass lead to values closer to 0.\hfill\break\indent
Finally I considered the case $v_2=10\cdot v_1$\ which lead
to a large enhancement of the standard QCD corrections due
to a $\tan\beta$\ dependence of the couplings. This decay
rate therefore may be a good decay mode to constrain the
 ratio of the vevs of the Higgs bosons once the
MSSM is proven to be the correct theory to describe nature.
\hfill\break\indent
Finally as it is well known I want to mention that the most competitive
decay mode to $t\rightarrow H^+b$\ is the equivalent decay mode
of $t\rightarrow W^+b$. Within the MSSM the electroweak corrections
to this decay mode was recently considered in [14]. The QCD
corrections with a gluino within the relevant loop diagram lead
to divergencies, which have to be renormalised. A full analysis
of the gluino contribution to this decay rate has not been done
yet and will be presented elsewhere [15].
\hfill\break\vskip.12cm\noindent
{\bf IV. ACKNOWLEDGMENT}\vskip.12cm
I would like to thank the physics department
of Carleton university for the use of their computer
facilities. The figures were done with the very user
friendly program PLOTDATA from TRIUMF.
\hfill\break\indent
This work was partially funded by funds from the N.S.E.R.C. of
Canada and les Fonds F.C.A.R. du Qu\'ebec.
\hfill\break\vskip.12cm\noindent
{\bf REFERENCES}\vskip.12cm
\item{[\ 1]}C.S. Li, B.Q. Hu and J.M. Yang, Phys.Rev.
{\bf D47}(1993)2865.
\item{[\ 2]}A. Czarnecki, Phys. Rev.{\bf D48}(1993)5250
\item{[\ 3]}C.S. Li and T.C. Yuan, Phys.Rev.{\bf D42}(1990)3088,
erratum-ibid{\bf D47}(1993)2556.
\item{[\ 4]}C.S. Li, Y.S Wei and J.M. Yang,
 Phys. Lett.{\bf B285}(1992)137.
\item{[\ 5]}J. Liu and Y.P. Yao, Phys.Rev.
{\bf D46}(1992)5196.
\item{[\ 6]}A. Czarnecki and S. Davidson, Phys.Rev.{\bf D48}
(1993)4183, Phys.Rev.{\bf D47}(1993)3063.
\item{[\ 7]}J.F. Gunion et al., "The Higgs Hunter's Guide"
(Addison-Wesley, Redwood City, CA, 1990).
\item{[\ 8]}H.P. Nilles, Phys.Rep.{\bf 110}(1984)1.
\item{[\ 9]}H.E. Haber and G.L. Kane, Phys.Rep.{\bf 117}(1985)75.
\item{[10]}A. Djouadi, M. Drees and H. K\"onig, Phys.Rev.
{\bf D48}(1993)3081.
\item{[11]} see e.g. HELIOS collaboration, T. Akesson et al,
Z.Phys.{\bf C52}(1991)219 and references therein.
\item{[12]}J. Ellis, D.V. Nanopoulos and D.A. Ross, Phys.Lett.
{\bf B305}(1993)375.
\item{[13]}CDF Collaboration, Fermilab preprint, April 1994.
\item{[14]}D. Garcia, R.A. Jim\'enez, J. Sol\`a and W. Hollik,
"Electroweak supersymmetric Quantum corrections to the top
quark decay", UAB-FT-323, hep-ph/9402341.
\item{[15]}H. K\"onig, in preparation.
\hfill\break\vskip.12cm\noindent
{\bf FIGURE CAPTIONS}\vskip.12cm
\item{Fig.1}The diagram with scalar quarks and gluino
within the loop, which contribute to the top quark decay
into a charged Higgs boson and bottom quark.
\item{Fig.2} The ratio of $\Gamma^1/\Gamma^0$\ as a function
of the scalar mass $m_S$\
for 3 different values
of the gluino mass: 3 GeV (solid line), 100 GeV (dotted line)
and 500 GeV (dash-dotted line) with $\mu=0=A_t$\ and $v_1=v_2$.
The top mass has been taken to be 174 GeV and the Higgs mass
to be $m_W$. The straight solid line is the standard QCD corrections
as given in [6].
\item{Fig.3}The same as in Fig.2 with $\mu=500$\ GeV and
$A_{\rm top}=m_S$.
\item{Fig.4}The same as in Fig.3 with $v_2=10\cdot v_1$.
\vfill\break
\end